\documentclass[11pt]{article}
\begin{document}

\def\be{\begin{equation}}
\def\ee{\end{equation}}
\def\ba{\begin{eqnarray}}
\def\ea{\end{eqnarray}}
\def\nn{\nonumber}
\def\lb{\label}
\def\dfrac{\displaystyle\frac}
\def\bb{\bibitem}
\def\E{{\cal E}}
\def\tJ{\tilde{J}}
\renewcommand{\theequation}{\arabic{section}.\arabic{equation}}
\begin{titlepage}
\date{}
\title{\begin{flushright}\begin{small}    LAPTH-064/22
\end{small} \end{flushright} \vspace{1.5cm}
Rotating traversable wormholes \\in Einstein-Maxwell theory}
\author{G\'erard Cl\'ement$^a$\thanks{Email: gclement@lapth.cnrs.fr},
Dmitri Gal'tsov$^{b,c}$\thanks{Email: galtsov@phys.msu.ru} \\ \\
$^a$ {\small LAPTh, Universit\'e Savoie Mont Blanc, CNRS, 9 chemin de Bellevue,} \\
{\small BP 110, F-74941 Annecy-le-Vieux cedex, France} \\
$^b$ {\small   Faculty of Physics, Moscow State University, 119899, Moscow, Russia }\\
$^c$ {\small  Kazan Federal University, 420008 Kazan, Russia}}

\maketitle

\begin{abstract}

It is well-known that traversable wormhole solutions to the Einstein equations require the existence of an exotic matter source violating the null energy condition. An apparent exception is the overcharged Kerr-Newman-NUT solution of the Einstein-Maxwell equations, which has all the properties of a geodesically complete, traversable wormhole spacetime. We show that the exotic matter sourcing this consists in two counter-rotating tensionless straight cosmic strings --- the Misner-Dirac strings --- which, as expected, violate the null energy condition. The wormhole possesses an ergoregion, but no superradiance. The geodesic motion in this spacetime is briefly discussed, and the absence of closed circular null geodesics is demonstrated.

\end{abstract}
\end{titlepage}
\setcounter{page}{2}
\setcounter{equation}{0}
\section{Introduction}
 In a search for theoretical models of traversable wormholes one simple possibility has remained unnoticed for a long time. Namely, the addition of a NUT parameter to some spherically symmetric solution with a naked singularity may transform it to a wormhole due to the replacement of the squared areal radius $r^2$ of the two-sphere with the quantity $r^2+n^2$.
In the previous article \cite{Clement:2015aka} we showed that the overextremal Reissner-Nordstr\"om-NUT solution \cite{brill} has all the properties of traversable wormholes \cite{wheeler, Morris:1988cz, Morris:1988tu} indeed (for a recent review see also \cite{Kundu:2021nwp}). 
This transformation of naked singularities into wormholes by the addition of a NUT charge extends to other theories \cite{Ayon-Beato:2015eca} and deserves further investigation, in particular by checking whether this change of the solution topology remains valid for rotating configurations. Here we show that the wormhole property is shared by the overcharged Kerr-Newman-NUT (KNN) solution, extending the theory to rotating wormholes, which have been discussed with some generality by Teo \cite{Teo:1998dp}, and more recently by Volkov \cite{volkov2021}. 

It is well known that traversable wormholes can only exist as solutions of the Einstein equations if they are supported by some exotic matter, violating the null energy condition 
(NEC). However the Maxwell stress-energy tensor of any regular field configuration satisfies the NEC \cite{Hawking:1973uf}, so the existence of NUT wormholes as solutions of the Einstein-Maxwell equations is puzzling. We shall show here that the solution to this puzzle is provided by the Misner strings associated with Taub-NUT and other NUTty solutions. To avoid closed timelike curves \cite{thorne} in all spacetime, we do not impose Misner's time periodicity condition \cite{Misner:1963fr}. The counterpart is that the Misner strings must be interpreted  as true spacetime singularities, as was first suggested by Bonnor \cite{bonnor}.  It was noticed in various contexts that Misner strings are supported by some distributional stress-energy tensor. In the black hole sector, the Komar mass and angular momentum  of Misner strings have been found to be non-zero \cite{Clement:2019ghi}. Though in the wormhole case the Komar integral for the string mass somewhat surprisingly vanishes, we will see that the Misner string stress-energy tensor does violates the NEC in the distributional sense. 

The KNN wormhole solution is presented in the next section. Geodesic motion in this spacetime is briefly discussed in Section 3. We prove that the null energy condition is violated in Section 4, and compute the Komar mass, Komar angular momentum and electric charge of the Misner strings in Section 5. Our results are briefly summarized in the concluding section.
 
\setcounter{equation}{0}
\section{The Kerr-Newman-NUT wormhole}

The metric of the  dyonic Kerr-Newman-NUT solution in Boyer-Lindquist coordinates can be conveniently written in two alternative forms
\ba
ds^2 &=&-\frac{\Delta}{\Sigma}(dt-\alpha d\varphi)^2+\Sigma\left( \frac{dr^2}{\Delta}+d\theta^2\right)+\frac{\sin^2\theta}{\Sigma}(\beta d\varphi - adt)^2 \lb{met1}\\
&=& - \frac{f}{\Sigma}(dt - \omega d\varphi)^2 + \Sigma\left[\frac{dr^2}{\Delta}   
 + d\theta^2 + \frac{\Delta}{f}\sin^2\theta d\varphi^2\right],\lb{met2}
\ea
where 
 \ba
 &&\alpha=a\sin^2\theta-2n\cos\theta,\quad\beta=r^2+n^2+a^2, \\&&\Delta =r^2-2mr + e^2 - n^2 + a^2,\\
&&
\Sigma =\beta - a\alpha = r^2 + (n + a\cos\theta)^2,\\  && f=\Delta - a^2\sin^2\theta, \quad\omega f = \alpha\Delta -a \beta\sin^2\theta,
\ea
where $e^2=q^2 + p^2$, with $q$ and $p$ the electric and magnetic charges, and $m$, $n$ and $a$ are the mass, NUT and rotational parameters. 

The outer root of the equation $\Delta=0$, $r_h=m+\sqrt{m^2+n^2-a^2-e^2}$
defines the location of the event horizon, which exists and is non-degenerate if $m^2+n^2-a^2-e^2>0$, exists and is degenerate (extremal) if $m^2+n^2-a^2-e^2=0$, and does not exist for 
\be\lb{over}
a^2 + e^2 - m^2 - n^2 \equiv b^2 > 0
\ee
(overspinning or overcharged solution). In the case (\ref{over}), the solution either has a naked ring singularity at $r=0, a\cos\theta =n$, if $|n|<a$ (we assume $a>0$), or is non-singular if 
\be\lb{wh}
|n| > a,
\ee
in which case the radial variable $r$ extends to the whole real axis, leading to a wormhole topology with two spacelike asymptotics $r = +\infty$ and $r = -\infty$. Note that the conditions (\ref{over}) and (\ref{wh}) together imply $e^2 = b^2 + m^2 + n^2 - a^2 > m^2$, so that the solution is overcharged rather than overspinning. For $a = 0$, this solution reduces to the non-rotating Reissner-Nordst\"om-NUT (or Brill) wormhole solution discussed in \cite{Clement:2015aka}. Note that the coordinates used in (\ref{met1}) or (\ref{met2}) are different from those used in Teo's metric ansatz \cite{Teo:1998dp}.

For the KNN wormhole the metric function 
\be
\Delta(r) = (r-m)^2 + b^2
\ee
 is always positive, so there is no horizon. The point $r=m$  corresponds to its minimal value $\Delta_0=b^2$. The metric function $f(r,\theta)$ is also positive definite for $b>a$. However, for 
 \be
 a>b,
 \ee
 the Killing vector $\partial_t$ becomes null on the boundary of the ergoregion $r<r_e$ where
\be
r_e=m+\sqrt{a^2\sin^2\theta-b^2}.
\ee
This boundary ends on non-zero value of the polar angle
\be
\pi-\theta_e>\theta>\theta_e,\qquad \theta_e=\arcsin\left(\frac{b}{a}\right).
\ee
Thus the ergoregion $r<r_e$ is bounded by the cones $\theta=\theta_e,\; \theta=\pi-\theta_e$. Its maximal radius at $\theta=\pi/2$ is 
\be
r_e^{\rm{max}}=m+\sqrt{a^2-b^2},
\ee
which exist if
\be
m^2+n^2-a^2<e^2<m^2+n^2.
\ee
Note that in the black hole case the ergoregion intersects the polar axis where it touches the horizon. So the absence of the horizon in the wormhole KNN sector correlates with non-intersection of the ergoregion boundary with the polar axis. The situation is similar to that of Teo's rotating wormhole example \cite{Teo:1998dp}. As was shown in \cite{Konoplya:2010kv}, ergoregions in wormholes do not imply superradiance (see also \cite{Mazza:2021rgq,Franzin:2022iai}).

The point $r=0$, where the metric function $\Sigma$ has the minimal $\theta$-dependent value $\Sigma_0=(n+a\cos\theta)^2>0 $, marks the location of the wormhole throat.  The second sheet of the wormhole is   described by the solution (\ref{met1}) in the range of the radial coordinate $-\infty<r<0$, which is asymptotically locally flat with a negative Schwarzschild mass $-m$, since
\be
g_{00}\sim 1+\frac{2m}{|r|}, \quad r\to -\infty.
\ee
Note that the global KNN wormhole metric is symmetric under the simultaneous reflections $r\to -r,\;m\to -m$, which becomes plane symmetry for $m=0$. It is also symmetric under the simultaneous reflections  $\theta \to \pi-\theta,\;n \to -n$.

On the wormhole throat $r=0$, the metric function $\omega$ is $\theta$-dependent. Thus, contrary to the case of the black hole horizon, the KNN wormhole throat is {\em differentially} rotating. Also, 
the metric function $\omega$ does not vanish for $\sin\theta = 0$, so that there are two distinct infinite Misner strings, North ($\theta = 0$) $S_+$, and South ($\theta = \pi$) $S_-$, rotating with opposite angular velocities
\be
\Omega_\pm = \frac1{\omega_\pm} = \mp \frac1{2n}.
\ee
The Misner strings are surrounded by the chronological horizon $g_{\varphi\varphi}=0$ inside which the $\varphi$-circles are closed timelike curves:
\be \beta^2\sin^2\theta < \alpha^2\Delta. \ee
This region spreads non-symmetrically through both sheets of the wormhole.

This metric is supported by the electromagnetic field with potential
\ba
A &=& v\,dt - \left(p\cos\theta + \alpha v\right)\,d\varphi, \nn\\
\Sigma v &=& - qr + p(n + a\cos\theta). \lb{pot}
\ea
Even if the magnetic charge $p$ vanishes, the azimuthal vector potential $A_\varphi$, does not vanish for $\sin\theta=0$, so that the two infinite Misner strings are also Dirac strings.

The magnetic field $B^r = F_{\theta\varphi}/\sqrt{|g|}$ takes on the wormhole throat $r=0$ the value
\be\lb{magneck}
B_0 = - \frac{p(n^2+a^2)}{(n + a\cos\theta)^4}.
\ee
In the case of
the purely magnetic wormhole ($q=0$), the largest throat magnetic field
\be
|B_{\rm max}| = \frac{|p|(n^2+a^2)}{(|n|-a)^4} > 
\dfrac{(|n|+a)^{1/2}(n^2+a^2)}{(|n|-a)^{7/2}}
\ee
is an increasing function of $a$. In the non-rotating case $a=0$, it was estimated \cite{Clement:2015aka} that a human being could traverse the wormhole throat without being harmed by the magnetic field if $|n| > 60$ lt-yr. This lower limit on $|n|$ would appear to be larger for a rotating magnetic wormhole. However, the throat magnetic field (\ref{magneck}) can be minimized for an axial geodesic. At either pole of the throat, the minimal magnetic field is achieved for $n \simeq \pm a$, leading to 
\be
|B_{\rm min}| = \frac{|p|}{8n^2} > \frac{|m|}{8n^2}
\ee
(from $p^2 = b^2 + m^2 + n^2 - a^2 > m^2$), which can be lowered to a comfortable value if the wormhole mass parameter $m$ is small before the NUT charge.

\setcounter{equation}{0}
\section{Geodesic motion}
\subsection{Geodesic equations}
The KNN metric admits two Killing vectors $k=\partial_t,\;m=\partial_\varphi$, ensuring conservation of the energy $E$ and the azimutal momentum $L$, and a second rank Killing tensor related to the Carter constant $K$. Normalizing the affine parameter on geodesics so that $g_{\mu\nu}{\dot x}^\mu {\dot x}^\nu =\epsilon$, where $\epsilon=-1,0$ for timelike and null respectively, one can obtain separated equations for radial and $\theta$-motion \cite{Cebeci:2015fie}:
\be\lb{r}
  \Sigma \frac{dr}{d\tau}=\pm\sqrt{R(r)},
\ee 
\be \lb{te}
\Sigma \frac{d\theta}{d\tau}=\pm\sqrt{\Theta(\theta)},
\ee
where the factor $\Sigma$ can be absorbed into redefinition of the parameter $d\tau\to\Sigma d\lambda$, and the potential terms read
\be\lb{rpot}
 R(r)=\left(\beta E -aL\right)^2 - \Delta(K - \epsilon r^2),
\ee
\be\lb{thpot}
 \Theta(\theta)=  K+\epsilon(n+a\cos\theta)^2-(L-\alpha E)^2/\sin^2\theta.
\ee

Once these equations are solved, the two other variables can be obtained by integrating the first order equations
\be\lb{time}
\Sigma\frac{dt}{d\tau}=\dfrac1{\Delta\sin^2\theta}\left[(L -\alpha E) \alpha\Delta + (\beta E - aL)\beta\sin^2\theta)\right],
\ee
\be\lb{varphi}
\Sigma\frac{d\varphi}{d\tau}=\dfrac1{\Delta\sin^2\theta}\left[(L -\alpha E) \Delta + (\beta E - aL)a\sin^2\theta)\right].
\ee 

The radial potential $R(r)$ is quartic in $r$, the coefficient of the leading term being $E^2+\epsilon$. It is clear that, whatever the wormhole parameter values, $R(r)$ will be positive definite for $E^2$ large enough, so that the wormhole is traversable. The different orbit types were discussed in the non-rotating case in \cite{Clement:2015aka}, and in the rotating case, including motion in the equatorial plane and spherical orbits of constant $r$, in \cite{Cebeci:2015fie} (see also \cite{Cebeci:2017xex,Mukherjee:2018dmm}), so we do not repeat this analysis here. 

\subsection{Angular motion}

The angular potential is of the form $\Theta(\theta) = \Xi(\xi)/\sin^2\theta$, where $\Xi(\xi)$ is quartic in $\xi=\cos\theta$. 
Depending on the values of the constants of motion, this may have several extrema, leading to oscillations between two turning points \cite{Cebeci:2015fie}. From (\ref{te}), it is clear that geodesics can cross the North (South) Misner strings only when the last term in the potential (\ref{thpot}) is non-singular at $\theta=0$ ($\theta=\pi$), which amounts to the  following relation between the energy and the azimuthal momentum
\be L=\mp 2nE.\ee
This condition coincides with that in the non-rotating case $a=0$ \cite{Clement:2015aka}. 

In that case, due to spherically symmetry, all angular trajectories were small circles on the spheres of constant $r$. For $a\neq0$, geodesic motion is only axisymmetric, but we still can solve the problem of motion with constant $\theta$ \cite{Cebeci:2015fie}. The Carter constant $K$ can be eliminated between the equations $\Theta(\theta)=0$ and $\Theta'(\theta)=0$ to yield the equation
 \be
(L-E\alpha)^2 + 2E(L-E\alpha)\dfrac{n+a\xi}{\xi}\sin^2\theta + a\dfrac{n+a\xi}{\xi}\sin^4\theta = 0,
 \ee
with $\xi \equiv \cos\theta$. Solving this for $$x \equiv (L-E\alpha)/\sin^2\theta$$ leads to the two roots
 \be
x_\pm = - \gamma\left(E \pm \sqrt{E^2 - a/\gamma}\right),
 \ee 
where $\gamma \equiv (n+a\xi)/\xi$. Thus, for given values of $E$ and $\theta_0$ ($0 \le \theta_0 \le \pi$), there are two geodesics winding around the North or South Misner string with the same energy $E$ and constant $\theta=\theta_0$, but two different values of the integration constants $K$ and $L$, as well as two different values of the angular velocity
 \be 
\frac{d\varphi_\pm}{d\tau}=\dfrac1{\Sigma}\left[x_\pm + \dfrac{a(\beta E - aL_\pm)}{\Delta}\right].
 \ee 
 
For $a=0$, the geodesics with $x=x_-=0$ and constant $\varphi$ describe the  motion of small extended particles infalling with the proper angular velocity (or spin) $2nE/\Sigma$ along any radial direction \cite{Clement:2015aka}. The same value of the spin for particles infalling along the North or South Misner string is obtained when one takes the limit $\sin\theta \to 0$ for the geodesics with $x=x_+$. 

For $a\neq0$, on the other hand, linear motion is only possible along the North or South Misner string. Setting $\theta=\dot{\theta}=0$ in (\ref{te}) one finds that this motion corresponds to
\be
L= \mp 2nE,\quad K=-\epsilon(n \pm a)^2.
\ee
Now, contrary to the non-rotating case, the two roots $x_\pm$ lead to two different limits for the spin of axially falling particles, which is consequently not well defined. 

\subsection{Chronology protection}

Closed timelike curves, which occur in a neighborhood of the Misner strings, are usually considered to violate causality. We have argued elsewhere \cite{Clement:2015cxa} that freely falling observers could encounter violations of causality only if there were closed timelike geodesics (CTGs), which were shown to be absent in the case of the non-rotating Reissner-Nordstr\"om-NUT solution. In the rotating case, calculations are more complicated, and we leave the general proof of the absence of CTGs for further work. As a first step towards this goal, let us show here that null geodesics with circular orbits cannot be closed, which presumably also implies the absence of CTGs with circular orbits.

Circular orbits have constant $r$ and $\theta$, so that the equations $R(r)=0$ and $\Theta(\theta)=0$ must be satisfied. Using these with $\epsilon=0$ in (\ref{time}) and (\ref{varphi}), we obtain
\be
\dfrac{dt}{d\tau} = \dfrac{KL}{(\beta E-aL)(L-\alpha E)}, \quad
\dfrac{d\varphi}{d\tau} = \dfrac{KE}{(\beta E-aL)(L-\alpha E)}.
\ee
From $\Theta'(\theta)=0$ with $\epsilon=0$, we obtain the relation between 
$L$, $E$ and $\theta$
\be
L = -E\left[\dfrac{2n}{\cos\theta} + a \sin^2\theta\right],
\ee
which can vanish only if 
\be\lb{cng}
0 < |n| < a/3\sqrt3.
\ee
However this parameter range is excluded by the regularity condition (\ref{wh}), so that $dt/d\varphi$ has a constant non-zero value along the geodesic, which cannot be closed. Note the role played by the wormhole regularity condition in proving this result, which might not hold for Kerr-Newman-Taub-NUT black holes in the parameter range (\ref{cng}).

\setcounter{equation}{0}
\section{Null energy condition}

As shown in \cite{Morris:1988tu}, the existence of static, spherically symmetric traversable wormhole configurations requires violation of the null energy condition
\be
R_{\mu\nu}k^\mu k^\nu \ge 0
\ee
for some null vector field $k(x)$ ($g_{\mu\nu}k^\mu k^\nu = 0$). This analysis was generalized by Teo \cite{Teo:1998dp} to the case of rotating wormholes. Teo showed in substance that the existence of stationary, axisymmetric traversable wormholes required again violation of the null energy condition. However Teo assumed in his proof that the various metric functions were regular on the axis, which is not the case of the function $\omega(x)$ of the KNN wormhole. While any solution to the Einstein-Maxwell equations is well-known \cite{Hawking:1973uf} to satisfy the null energy condition, we check here directly that it is satisfied by the KNN solution everywhere except for Misner string singularities.

\subsection{Checking the NEC in the bulk}

Consider the bulk contribution (outside the Misner strings) to the KNN Ricci tensor, with the only non-vanishing mixed elements
\be 
 R^\varphi_\varphi=-R^t_t=\frac{e^2}{\Sigma^3}\left(a\alpha+\beta\right),\quad 
 R^\theta_\theta=-R^r_r=\frac{e^2}{\Sigma^2},\ee 
\be R^t_\varphi=\frac{2 e^2\alpha\beta}{\Sigma^3},\quad R_t^\varphi=-\frac{2a e^2}{\Sigma^3},
\ee
with $e^2 = q^2 + p^2$. The square of the Ricci tensor has a very simple form \be
R_{\mu\nu}R^{\mu\nu}=\frac{4 e^4}{\Sigma^4}.
\ee

To prove the validity of the NEC with respect to the bulk fields we have to verify positive definiteness of the quadratic form obtained with account for $k_\mu k^\mu=0$:
\ba
 R^\mu_\nu k^\nu k_\mu &=& \dfrac{2e^2}{\Sigma^3}\left[\Sigma k^\theta k_\theta   +(\beta k^\varphi-a k^t)(k_\varphi+\alpha k_t) \right]\nn \\
 &=& \dfrac{2e^2}{\Sigma}\left[(k^\theta)^2 + \Sigma^{-2}\sin^2\theta(\beta k^\varphi - a k^t)^2 \right],\lb{nullbulk}
\ea
where we have used the form (\ref{met1}) of the metric. 

\subsection{NEC violation by Misner strings} 

Besides this non-negative bulk contribution, we must also take into account the Misner strings, tensionless rotating cosmic strings, which generate a distributional contribution to the Ricci tensor \cite{djh,Clement:1985nk}. Recall the Israel-Wilson  dimensional reduction of the Einstein-Maxwell system \cite{Israel:1972vx} (for derivation in the current notation see, e.g., \cite{Galtsov:1995mb}). Using the ansatz for stationary four-dimensional metrics
\be ds^2 = -F\left(dt -\omega_i dx^i \right)^2 + F^{-1} h_{ij}dx^i
dx^j \ee
($F = f/\Sigma$ in the present case), one can express the mixed components of the Ricci tensor as
\be 
R^i_t =\frac{F}{2\sqrt{h}}\epsilon^{ijk}\tau_{k,j}, \lb{mixRicci}
\ee
where the twist $\tau_{k}$ is the three-dimensional dual to the rotation
 two-form $d\omega$,
\be
\tau^i = \epsilon^{ijk}\frac{F^2}{\sqrt{h}}\partial_j\omega_k.
\ee 
Near a Misner string $u \equiv \sin\theta=0$, we have
\be
h \simeq dr^2 + \Delta(du^2 + u^2 d\varphi^2), \quad \omega \simeq \mp 2n d\varphi.
\ee
Transforming to local coordinates 
 $x^1 = r$, $x^2 = u\cos\varphi$, $x^3 = u\sin\varphi$ leads to
 \be 
h \simeq (dx^1)^2 + \Delta\left[ (dx^2)^2 + (dx^3)^2\right], \quad \omega \simeq \pm 2n \epsilon_{1ij} \partial_j\ln{u} dx^i 
 \ee
($u^2 = (x^2)^2 + (x^3)^2$). Accordingly,
 \be\lb{tausing}
\tau_1 = \tau^1 \simeq \mp \dfrac{2n\Delta}{\Sigma^2}\,\nabla^2\ln{u} =
\mp \dfrac{4\pi n\Delta}{\Sigma^2}\,\delta^2(x)
 \ee
(the Dirac delta distribution in the two-plane $(x^2,x^3)$ perpendicular to the Misner string). 
We then obtain from (\ref{mixRicci}) the North Misner string contribution to the mixed Ricci tensor density:
\be
\sqrt{|g|}(R_S)_t^i = \mp\dfrac{2\pi n\Delta}{\Sigma^2} \epsilon^{1ij} \partial_j \delta^2(x).
 \ee 
Integrating over all space,
\be
X_S(r) \equiv \int\sqrt{|g|}(R_S)_t^i k^t k_i d^3x = \pm 2\pi n \epsilon^{1ij} \int_{-\infty}^{+\infty}\partial_j\left(\dfrac{\Delta}{\Sigma^2} k^t k_i\right)_S\,dr,
 \ee 
where the partial derivatives must be evaluated on the Misner string $x^2 = x^3 = 0$. 

Choose the null vector field
 \be\lb{testnull}
k^t = \lambda\beta\sin\theta, \; k^r = \varepsilon\lambda\Delta\sin\theta, \;
k^\theta = 0, \; k^\varphi =\lambda a\sin\theta
\ee
($\varepsilon = \pm1$), with $\lambda(r,\theta)$ an arbitrary function.
This implies for $i=2,3$:
\be
k_i = \dfrac{\lambda\alpha\Delta}{\sin\theta}\epsilon_{1ij}x^j.
\ee
The bulk contribution (\ref{nullbulk}) vanishes identically, while the Misner string contribution
\be
X_S(r) = - 8\pi n^2\int_{-\infty}^{+\infty}\dfrac{{\lambda_S}^2\Delta^2\beta}
{{\Sigma_S}^2}\,dr
\ee
is negative definite.

Note that the twist $\tau^1$ induces also the following singular contributions to the four-dimensional Ricci-tensor:
\be\lb{tau2}
(R_{tt})_{\rm twist}=\frac12 (\tau^1)^2,\quad (R^{ij})_{\rm twist}=-\frac12 h^{ij} (\tau^1)^2 \quad (i,j=2,3).
\ee
Substituting $\tau^1 = \tau^1_{\rm reg} + \tau^1_{\rm sing}$, where the regular
part is finite, and the singular part is given by (\ref{tausing}), we see that the $(\tau^1)^2$ contributions to the Ricci tensor (\ref{tau2}) will lead to additional $\delta^2(x)$ contributions. It is not necessary to compute these explicitly, as the factor of $(\tau^1)^2$ in the product $(R_{tt})_{\rm twist}(k^t)^2$ is proportional to $\sin^2\theta$ for the vector field (\ref{testnull}), and so does not contribute, while that in the product $(R^{ij})_{\rm twist} k_ik_j$ is negative definite, and so strengthens our result. The conclusion is that the KNN metric (\ref{met1}) does violate the null energy condition.

\setcounter{equation}{0}
\section{Komar charges}
%%%%%%%%%%%%%%%%%%%%%%%%%%%%%%%%%
Now we inquire what are the physical characteristics of the exotic matter sourcing the KNN wormhole --- the Misner strings. For this purpose, let us first recall the Weyl form of the metric for stationary axisymmetric spacetimes,
\begin{equation}                    \label{Gzr}
 ds^2 = G_{ab}(\rho,z) dx^a dx^b + e^{2\nu(\rho,z)}(d\rho^2 +  dz^2)\,,
\end{equation}
($x^a=t,\varphi$), with $\rho = \sqrt{|\det G|}$. Killing horizons  corrrespond to (finite or infinite) segments (``rods'') of the $z$-axis where $G$ is singular, $\rho=0$. Tomimatsu designed a scheme \cite{Tomimatsu:1983qc,Clement:2017otx} to compute the Komar mass and Komar angular momentum \cite{komar}, as well as the electric charge carried by black hole horizons (timelike Killing horizons). This was extended in \cite{Clement:2019ghi} to Misner strings, which are Killing horizons for the spacelike Killing vectors $\xi_\pm = \partial_t + \Omega_\pm\partial_\varphi$. 

In the case of systems of black holes and Misner strings, the total electric charge observed at infinity 
\be 
Q_\infty = \frac1{4\pi}\oint_ {\Sigma_{\infty}}F^{\mu\nu} d\Sigma_{\mu\nu} \lb{q}
\ee 
($F^{\mu\nu}$ being the Maxwell tensor) is equal, by the Gauss theorem, to the sum of the electric charges $Q_n$ carried by the various inner boundaries $\Sigma_n$ (Killing horizons), $Q_\infty = \sum_n Q_n$. In the present wormhole case, one must take care that the hypersurface at spacelike infinity enclosing the sources has two components $\Sigma_\infty^\pm$ ($r\to\pm\infty$), so that the Gauss theorem leads to
\be
Q_\infty^+ + Q_\infty^- = \sum_n Q_n,
\ee 
where the value of the left-hand side is $q -q = 0$, while (in the absence of black hole horizons) the sum on the right-hand side contains only two terms $Q_\pm$ corresponding to the two infinite Misner strings. 

Similarly, the Komar mass and angular momentum observed at either end of the wormhole are defined as the surface integrals:
\ba
 &&M_\infty^\pm = \frac1{4\pi}\oint_ {\Sigma_{\pm\infty}}D^\nu k^{\mu}d\Sigma_{\mu\nu}, \lb{km} \\
&&J_\infty^\pm = -\frac1{8\pi}\oint_{\Sigma_{\pm\infty}}D^\nu m^{\mu}d\Sigma_{\mu\nu}, \lb{kj}
\ea
where $k^\mu = \delta^\mu_t$ and $m^\mu = \delta^\mu_\varphi$ are
the Killing vectors associated with time translations and rotations
around the North or South pole, and $D^\nu$ is the covariant derivative. By using the Einstein equations and the three-dimensional Ostrogradsky theorem, one arrives at the balance equations
 \be
M_\infty^+ + M_\infty^- = \sum_n M_n, \qquad J_\infty^+ + J_\infty^- = \sum_n J_n,
 \ee
where again the left-hand sides have the values $m-m=0$ and $am-am=0$, and the Komar charges $M_n$ and $J_n$ of the Killing horizons (the two Misner strings) are given in terms of the metric and electromagnetic fields by equations (3.7) and (3.8) of \cite{Clement:2019ghi}.

The final step in the Tomimatsu scheme involves transforming the expressions of the Killing horizon charges into ``mixed'' formulas involving both the fields and the gravitational and electromagnetic Ernst potentials $\E$ and $\psi$:
\ba 
Q_n &=& \frac{\omega_n}2{\rm Im}\psi \Big {\vert}_{a_n}^{b_n}, \\
M_n &=& \left[\frac{\omega_n}{4}{\rm Im}\E +\frac12 (A_\varphi\,{\rm Im}\psi)\right]_{a_n}^{b_n}, \\
J_n &=& \frac{\omega_n}{4}  \left\{ a_n - b_n + \left[\omega_n\left( {\rm Im}\E /2-\Phi_n {\rm Im}\psi\right)+  A_\varphi\,{\rm Im}\psi \right] \Big {\vert}_{a_n}^{b_n}\right\},
\ea 
where $\omega_n= 1/\Omega_n$ is the constant value of $\omega$ on the Killing horizon, $-\Phi_n= v +\Omega_n A_\varphi$ is the constant rod electric potential in the co-rotating frame, and $a_n$ and $b_n$ are the $z$ coordinates of the origin and end of the rod $n$.

In the present case, the Ernst potentials for the KNN solution are
\be
\E = \frac{r + i(a\cos\theta-n)}{r -2m+ i(a\cos\theta+n)}, \qquad \psi = \frac{-q+ip}{r+ i(a\cos\theta+n)},
\ee
while $a_\pm = -\infty$, $b_\pm = +\infty$.
Since ${\rm Im}\E$ and ${\rm Im}\psi$ vanish at $\pm\infty$, while $A_\varphi$ stays finite, the two Misner strings are chargeless and massless,
\be
Q_\pm = 0, \qquad M_\pm = 0.
\ee
On the other hand, the Komar angular momenta take opposite diverging values,
\be 
J_\pm = \pm nR,
\ee
where $R$ is an infrared cutoff. These are the values which are expected for two counter-rotating, tensionless and chargeless straight cosmic strings.

\section {Conclusion}

Continuing our previous analysis of the overcharged Reissner-Nordstr\"om-NUT metric, we have shown that the rotating Kerr-Newman-NUT spacetime is also a geodesic complete, traversable wormhole in the superextremal case. However this is puzzling, as it is known that the existence of rotating or non-rotating traversable wormhole solutions to the Einstein equations requires violation of the null energy condition, which is satisfied in the Einstein-Maxwell theory. We have solved this puzzle by showing that although the NEC is perfectly satisfied in the bulk, it is violated by the Misner strings, which appear as additional matter distributional singularities in the Bonnor interpretation adopted here.

We conclude that these NUT wormholes are created by an exotic (i.e. NEC-violating) matter, different from those, such as thin-shell wormholes, envisioned so far. This exotic matter consists in two counter-rotating and tensionless  straight cosmic strings --- the Misner-Dirac strings. The analysis of geodesic motion shows that, as in the case of the rotating wormholes discussed in \cite{Teo:1998dp}, most timelike or null geodesics which traverse the wormhole from one end to the other avoid altogether this exotic matter. Moreover, those geodesics which do cross one or the other Misner string do it smoothly.   

An unpleasant feature of NUTty spacetimes is the presence of closed timelike curves surrounding the Misner strings. But it was argued \cite{Clement:2015cxa,Clement:2016mll} that these curves cannot be associated with true physical movements and are therefore classically harmless. This point of view is attracting recently more and more attention and stimulates further research \cite{Barrientos:2022avi}. Closed timelike or null geodesics, which could lead to observable violations of causality, were shown to be absent in the non-rotating case \cite{Clement:2015aka}. Here we have shown in a first step that null geodesics with circular orbits cannot be closed. We leave for future work the more involved analysis of the possibility of observing causality violation in geodesic as well as in charged particle motion \cite{Clement:2016mll}.

There still remains the problem concerning wave propagation in the framework of field theories, since rejection of the Misner periodicity leads to a singularity of the field modes on the polar axis. But this is exactly what can be expected in a singular space-time, so this is part of a wider problem (not yet solved)  of interpretation of space-time singularities. This problem is also closely related to the stability issue, depending on the choice of boundary conditions for spacetime perturbations near a singularity \cite{Gibbons:2004au,Sadhu:2012ur}.

Finally, we note that overcharged solutions can hardly be formed in the gravitational collapse in realistic astrophysical situations, but their possible existence as topological defects created in the early universe is worth to be explored.

\section*{Acknowledgments}
The work was partially supported by the Russian Foundation for Basic Research on the project 20-52-18012 Bulg-a, by the Scientific and Educational School of Moscow State University “Fundamental and Applied Space Research”, and the Strategic Academic Leadership Program ``Priority 2030'' of the
Kazan Federal University.

\end{document}